\documentclass[12pt,letterpaper]{article}
 \usepackage{citesort}
 \usepackage{amsfonts}
 \usepackage{amssymb}
 \usepackage[dvips]{graphicx}
 \usepackage{latexsym}
 \newcommand{\beq}{\begin{equation}}
 \newcommand{\eeq}{\end{equation}}
 \newcommand{\bea}{\begin{eqnarray}}
 \newcommand{\eea}{\end{eqnarray}}
 \newcommand{\lpart}{\raise.3ex\hbox{$\stackrel{\leftarrow}{\partial}$}}
 \newcommand{\rpart}{\raise.3ex\hbox{$\stackrel{\rightarrow}{\partial}$}}
 \newcommand{\ldr}{\raise.3ex\hbox{$\stackrel{\leftarrow}{\delta^r}$}}

 \newcommand{\al}{\alpha}
 \newcommand{\be}{\beta}

\begin{document}

\begin{flushright}
hep-ph/0208241\\
HIP-2002-37/TH \\
August 27, 2002\\
\end{flushright}
\vspace*{3mm}
\begin{center}
{\Large {\bf Superhorizon Curvaton Amplitude in Inflation} \\[.1cm]
{\bf and Pre-Big Bang Cosmology}\\}
\vspace*{12mm}
{\bf Martin S. Sloth\footnote{\tt e-mail: martin.sloth@helsinki.fi}\\}
\vspace{3mm}
{\it Helsinki Institute of Physics\\[.1cm]
P.O. Box 64, FIN-00014 University of Helsinki, Finland}
\vspace{1cm}
\begin{abstract}
\noindent We follow the evolution of the curvaton on superhorizon
scales and check that the spectral tilt of the
curvaton perturbations is unchanged as the curvaton becomes
non-relativistic. Both inflation and pre-big bang
cosmology can be treated since the curvaton mechanism within the two scenarios
works the same way. We also discuss the amplitude of the
density perturbations, which leads to some interesting constrains on
the pre-big bang scenario. It is shown that within a $SL(3,\mathcal{R})$ 
non-linear sigma model one of the three axions has the right coupling to 
the dilaton and moduli to yield a flat spectrum with a high string scale, 
if a quadratic non-perturbative potential is generated and an 
intermediate string phase lasts long enough.
\end{abstract}
\end{center}
\setcounter{footnote}{0}
\baselineskip16pt

\section{Introduction}

The curvaton mechanism is an alternative way to generate the
initial adiabatic density perturbations leading to the observed
CMB anisotropies. It has recently generated some interest both in
the context of pre-big bang (PBB) cosmology
\cite{Enqvist:2001zp,Bozza:2002fp,Gasperini:2002bn,Tsujikawa:2001ad,Notari:2002yc}
and inflation
\cite{Lyth:2001nq,Lyth:2002my,Moroi:2001ct,Bartolo:2002vf,Moroi:2002rd}.
The mechanism was first discussed more than ten years ago in
\cite{Mollerach:hu}, but also later briefly in
\cite{Linde:1996gt}. The curvaton is an effectively massive
scalar particle that only gets to dominate energy density after
the beginning of the first radiation dominated era. As it decays,
the initial isocurvature perturbations of the curvaton is
converted into adiabatic ones. The name ``curvaton'' was first
suggested in \cite{Lyth:2001nq}. In inflation models the
mechanism has received interest as an alternative to the usual
scenario where the density perturbations are created by the
quantum fluctuation of the inflaton, which can lead to a
significant component of both isocurvature modes and non-Gaussian
fluctuations. Within the PBB scenario, the mechanism is even more
interesting since it might be needed for the scenario to yield
the observed flat spectrum of adiabatic density perturbations.

It has been known for some time that the adiabatic perturbations
created in the PBB phase has strong spectral tilt $\Delta n=3$
where $\Delta n=0$ is scale invariant \cite{Brustein:1994kn} unless an
exponential potential for the dilaton is included
\cite{Finelli:2001sr,Finelli:2002we}. Non standard matching conditions like those used 
in the ekpyrotic scenario has been applied also to PBB by Durrer and Vernizzi
\cite{Durrer:2002jn}, who find $\Delta n=-1$, although there are 
some arguments that the original matching conditions are correct
\cite{Cartier:2001is,Tsujikawa:2002qc}. However, a flat spectrum
of isocurvature perturbations is generated by the axion fields,
which do not contribute to the background in the PBB dynamics.
Since the axion spectrum is flat, as opposed to the very blue
spectrum of the dilaton, they will dominate the energy density
fluctuation spectrum. By means of the curvaton mechanism, those
isocurvature density fluctuations can be converted into adiabatic
ones.

In this paper we shall concentrate on some aspects of the
curvaton mechanism, which we believe have not been treated in
detail before. It can be argued very simply \cite{Lyth:2001nq},
that any evolution of the curvaton field perturbation on
superhorizon scales only leads to an overall scale independent
factor, which will conserve the flatness of the spectrum. Here we
confirm the validity of the argument as we follow the evolution
of the curvaton on superhorizon scales till after the modes has
become non-relativistic. We show explicitly that the spectrum
stays flat after the curvaton starts to oscillate in its
quadratic potential. The argument is similar to work done in
connection to the ``seed mechanism'' in \cite{Durrer:1998sv}.

It has also been noted that if the curvaton of the PBB scenario is
the NS-NS axion, then with a quadratic potential, the field
fluctuations of the curvaton seems to be too large to be
compatible with a high string scale of order $M_s\sim10^{-2}m_p$
\cite{Enqvist:2001zp,Bozza:2002fp}. It was first suggested that a
periodic potential usually expected for the axion could damp the
fluctuation to the right level \cite{Enqvist:2001zp}, but
it has also been suggested that a break in the spectrum can do the
job \cite{Bozza:2002fp}. We will investigate this further and show
that an intermediate string phase can help provided the coupling of 
the axion to the dilaton and the moduli satisfy certain requirements.
We give an explicit example where the requirements are satisfied.

\section{The curvaton model}

During the inflationary or PBB era the curvaton field is constant and does not
contribute to the background. However, the canonically normalized
quantum fluctuation of the curvaton is amplified. In this epoch the curvaton
field contributes to the total energy density like a
cosmological constant. After the universe has entered the first
radiation dominated era at conformal time $\eta=\eta_r$ and when the
Hubble rate subsequently reaches $H_{osc}\simeq m$, the curvaton
starts to oscillate in its potential. From this point on, the
energy density of the curvaton will behave like matter and falls
off like $\rho_{\sigma}\propto a^{-3}$. This implies that the
curvaton will soon dominate energy density and at this point the
Hubble rate is $H_{dom}\sim \Omega m\sigma_r^4$. Here $\Omega$ is
a conformal factor relevant only in the PBB scenario and
$\sigma_r$ is the value of the curvaton field at the beginning of
the first radiation dominated epoch. Finally the curvaton decays
into radiation, and the curvaton fluctuations induces the initial
adiabatic density perturbations in the cosmic fluid leading to the
CMB anisotropies.

From requiring that the curvaton only dominates energy density
after it has started to oscillate in its potential\footnote{If
this is not the case, we are not really considering a curvaton
model, since the curvaton will then lead to some additional
inflation \cite{Bozza:2002fp}.} we must constrain ourselves to
$\sigma_r/m_p<1/\Omega$ ($m_p$ is the Planck mass).

Since the curvaton models within the PBB and the inflationary
scenarios are very similar, we will treat them both. In this
section we will set up the notation and, as we will confine
ourself to a curvaton with a quadratic potential, parameterize
the action for the curvaton $\sigma$ in the following way
 \beq
S=\frac{1}{2}\int
d^4x\sqrt{-g}e^{l\phi}e^{mb}\left[\left(\nabla\sigma\right)^2+m^2\sigma^2\right]~.
 \eeq
Where $g$ is the determinant of the metric, $b$ is the moduli of
the internal dimension and $\phi$ is the dilaton. The
inflationary case is obtained by taking $l=m=0$, while in the PBB
set-up $l,m$ can in principle take any positive or negative value of
order one.

We will be interested in the inhomogeneous fluctuations of the
curvaton field around a homogeneous background field
$\sigma= \left<\sigma\right>+\delta\sigma$. In the following $\sigma$
denotes only the background value of the curvaton
field. It is useful to define a pump field
$S=a\exp(l\phi/2+mb/2)$. Then the canonically normalized
perturbation field is $\psi_k=S\delta \sigma(k)$ and the
perturbation equation can be written as
 \beq \label{peq}
\psi_k''+(k^2-\frac{S''}{S}+a^2m^2)\psi_k=0~.
 \eeq
Here $'$ denotes derivative with respect to the conformal time
$\eta$. In accordance with the curvaton scenario, we will assume
that the curvaton does not affect the background evolution until
after the field has started to oscillate in its potential. 
We will also assume that the curvature perturbations at the beginning 
of the first radiation dominated era are negligible, so we can ignore 
the coupling to other perturbations at least until the curvaton 
dominates energy density.

Below we will remind about some well known features of inflation and PBB,
paving the way for the discussion in the following sections.

\subsection{Inflation}

In an inflationary set-up the pump field is just $S=a$. During
inflation, when the mass of the curvaton can be neglected\footnote{The mass 
can be neglected until at least a few Hubble times after Horizon exit,
where the curvaton fluctuations are frozen.}, the
perturbation equation becomes
 \beq
\psi_k''+(k^2-\frac{a''}{a})\psi_k=0~.
 \eeq
In case of de Sitter inflation we have $a''/a=2/\eta^2$. If we
define $\lambda(\lambda-1)/\eta^2\equiv a''/a$, then the solution
to the perturbation equation normalized to a vacuum fluctuation in
the infinite past is
 \beq
\psi_k=\frac{\sqrt{\pi}}{2}\sqrt{\eta}H_{\mu}^{(2)}\left(|k\eta|\right)~~,\qquad
\mu=|\lambda-1/2|~.
 \eeq
Thus, for de Sitter inflation $\mu=3/2$. $H_{\mu}^{(2)}$ is the
Hankel function of the second kind. At horizon exit
$k=a_*H_*=1/\eta_*$ and with $\mu=3/2$ we obtain, by using the
small argument expansion of the Hankel function,
 \beq \label{pspec0.1}
\mathcal{P}_{\delta\sigma}^{1/2}(k)
=\frac{k^{3/2}}{\sqrt{2}\pi}|\delta\sigma(k)|=\frac{1}{2\pi}H_*~~,\qquad
\mu=3/2
 \eeq
where $H_*$ is the Hubble parameter at horizon exit. Since $H$ is
approximately constant during inflation $H_*$ is equal to the Hubble
parameter at the beginning of the first radiation dominated epoch
$H_r$. The power-spectrum in equation (\ref{pspec0.1}) was applied in
\cite{Lyth:2001nq}.

\subsection{Pre-Big Bang}
The PBB scenario \cite{Veneziano:1991ek,Gasperini:1993hu,Gasperini:1992em}
is based on a four-dimensional effective tree-level string theory action,
which in Einstein frame can be written as
 \beq
S_{eff}=\frac{1}{2\kappa^2}\int
d^4x\sqrt{-g}\left[\mathcal{R}+\frac{1}{2}(\nabla\phi)^2-\frac{1}{2}(\nabla b)^2+\mathcal{L}_{matter}\right]~,
 \eeq
where $\kappa^2=8\pi/M_s^2$ and $M_s$ is the string mass.
Here $\phi$ and $b$ are the universal four-dimensional moduli
fields. This action can be derived from 10-dimensional string
theory compactified on some 6-dimensional compact internal manifold.
The matter Lagrangian, $\mathcal{L}_{matter}$, is composed of
gauge fields and axions, which we assume do not to contribute to the background.

The solution in Einstein frame to the equations of motion for the 
background fields can in the spatially flat case be parameterized
as \cite{Lidsey:1999mc}
 \beq
a=a_r\left|\frac{\eta}{\eta_r}\right|^{1/2}\qquad,\qquad
e^{\phi}=e^{\phi_r}\left|\frac{\eta}{\eta_r}\right|^{\sqrt{3}\cos\zeta}\qquad,\qquad
e^{b}=e^{b_r}\left|\frac{\eta}{\eta_r}\right|^{\sqrt{3}\sin\zeta}~.
 \eeq

As mentioned in the introduction, in the PBB scenario we identify the 
curvaton with the axion field. Let us split out the part of the matter 
Lagrangian that contains the axion field
$\mathcal{L}_{matter}=\mathcal{L}_{gauge}+\mathcal{L}_{axion}$,
where we will assume that the axion Lagrangian has the general
form
 \beq
\int d^4 x\sqrt{-g}\mathcal{L}_{axion}=\frac{1}{2} \int d\eta d^3
x a^{4}e^{l\phi}e^{mb}\left[(\nabla
\sigma)^2+m^2\sigma^2\right]~.
 \eeq
In the PBB era we can (like in the inflationary phase) ignore the
mass term in the resulting perturbation equation and the solution
to the perturbation equation normalized to a vacuum fluctuation in
the infinite past is like in the previous subsection
 \beq
\psi_k=\frac{\sqrt{\pi}}{2}\sqrt{\eta}H_{\mu}^{(2)}
\left(|k\eta|\right)~~,\qquad \mu=|\lambda-1/2|
 \eeq
with $\lambda(\lambda-1)/\eta^2=S''/S$. The only difference being that 
$\mu$ now depends on the PBB background dynamics through the non-trivial pump field
$S=a\exp(l\phi/2+mb/2)$ and in principle can take any value
depending on the values of $l$, $m$ and $\zeta$.

Having set up the notation, in such a way that the curvaton perturbation
spectrum takes the same form in the inflationary and PBB scenarios,
we are ready to discuss the superhorizon curvaton amplitude during
the first radiation dominated epoch.

\section{Superhorizon curvaton amplitude}

In this section we calculate the amplitude of the curvaton
perturbation outside the horizon in the non-relativistic regime
by matching its solution to the spectrum obtained at the onset of
the first radiation dominated epoch.

Assume that the perturbation spectrum of the curvaton at the
beginning of the first radiation dominated phase is given by
 \beq \label{pspec}
\psi_k=\frac{\sqrt{\pi}}{2}\sqrt{\eta}H_{\mu}^{(2)}(|k\eta|)~.
 \eeq

Let $\eta_r$ denote the beginning of the radiation dominated
phase. In the radiation dominated era $\eta>\eta_r$, the scale
factor is proportional to the conformal time $a\sim \eta$. Since
at this point in PBB cosmology the dilaton and moduli fields are frozen
to a constant value, then for both inflation and PBB
cosmology we can ignore the effective potential term $S''/S$ in
the perturbation equation (\ref{peq}), and it simply reads
 \beq \label{peq1}
\psi_k''+(k^2+a^2m^2)\psi_k=0~~,\qquad \eta>\eta_r~.
 \eeq

Following \cite{Durrer:1998sv} and defining $\al = mH_ra_r^2$, where
$a=a_r\eta/\eta_r$ and $H_r=1/(a_r\eta_r)$, it is useful to note
that $\al^2\eta^2=a^2m^2$, such that the equation (\ref{peq1})
becomes
 \beq \label{peq2}
\psi_k''+(k^2+\al^2\eta^2)\psi_k=0~~,\qquad \eta>\eta_r~.
 \eeq
Finally, by defining $x=\eta(2\al)^{1/2}$ and $-b=k^2/(2\al)$
one obtains
 \beq
\frac{d^2\psi_k}{dx^2}+\left(\frac{x^2}{4}-b\right)\psi_k=0~.
 \eeq
This equation can be solved exactly by parabolic cylinder functions
 \beq \label{pspec1}
\psi_k=Ay_1(b,x)+By_2(b,x)
 \eeq
where $y_1$ and $y_2$ are the even/odd parts of the parabolic
cylinder functions.

We are now in position to match (\ref{pspec})
and (\ref{pspec1}) at the beginning of the radiation dominated
era $\eta=\eta_r$. Let $H_m=1/(a_m\eta_m)\sim m$, then
$k_m=1/\eta_m$ marks the scale that re-enters just as it becomes
non-relativistic. At the matching time $x\sim \eta/\eta_m<1$ and
for modes outside the horizon $x^2b<<1$, we can use the expansion
 \bea
y_1(b,x)&=&1+b\frac{x^2}{2!}+\dots \nonumber\\
y_2(b,x)&=&x+b\frac{x^2}{2!}+\dots
 \eea
This implies \cite{Durrer:1998sv}
 \beq \label{pspec1.2}
\psi_k = C(\mu)(2\eta_r\al)^{-1/2}|k\eta_r|^{-\mu}y_2(b,x)~.
 \eeq
with
 \beq
C(\mu)=\sqrt{2}\frac{\Gamma(\mu)2^{\mu}}{\Gamma(3/2)2^{3/2}}~.
 \eeq
The curvaton perturbation spectrum in equation (\ref{pspec1.2})
was also applied at the beginning of the radiation dominated era
in \cite{Bozza:2002fp}, where we can use $x<<1$ i.e. $y_2\sim x$
to yield
 \beq
\delta
\sigma(k)=C(\mu)a_r^{-3/2}H_r^{-1/2}\left(\frac{k}{k_r}\right)^{-\mu}~,\qquad \eta<\eta_{osc}
 \eeq
thus, before the curvaton starts to oscillate in its potential
$\sigma$ as well as $\delta\sigma(k)$ are constant as we expected.

But, as in \cite{Durrer:1998sv}, we can follow the superhorizon evolution
of the perturbation spectrum to the matter dominated era. After
the curvaton has started to oscillate in the potential $\eta>\eta_m$
i.e. $x>1$ we get
 \beq
\psi_k=
\frac{A}{(2\al\eta)^{1/2}}\left(\frac{k^2}{2\al}\right)^{1/4}|k\eta_r|^{-\mu-1/2}
\sin\left(\frac{m}{H}+\frac{1}{8}\pi\right)
 \eeq
where $A=\sqrt{2}C(\mu)\sqrt{\Gamma(1/4)/\Gamma(3/4)}$. By using
$k^2/\al=k^2H_r/(k_r^2m)$ we find straightforwardly that the amplitude 
of the curvaton oscillation is
 \beq \label{pspec2}
\delta \sigma(k)
=\frac{\tilde{A}}{a\sqrt{am}}\left(\frac{H_r}{m}\right)^{1/4}\left(\frac{k}{k_r}\right)^{-\mu}~~,
\qquad \eta>\eta_{osc}
 \eeq
where
 \beq
\tilde{A}=2^{-3/4}\Omega^{-1}A~.
 \eeq
For the inflationary scenario $\tilde{A}=2^{-3/4}A$, while for
PBB $\tilde{A}=2^{-3/4}\exp(-l\phi_r/2-mb_r/2)A$.

It is now easy to see that
 \beq
\delta \sigma(k)=\left(\frac{a_{osc}}{a}\right)^{3/2}\delta
\sigma_{osc}(k)~~,\qquad\eta>\eta_{osc}
 \eeq
and using the fact that the universe is radiation dominated until the curvaton starts to oscillate in its potential, we may use also $m=H_{osc}=H_r(a_r/a_{osc})^2$ and we obtain
 \beq
\delta
\sigma_{osc}(k)=\frac{\tilde{A}}{C(\mu)}\left(\frac{a_r}{a_{osc}}\right)^{3/2}\left(\frac{H_r}{m}\right)^{3/4}\delta
\sigma_{r}(k)=\frac{\tilde{A}}{C(\mu)}\delta
\sigma_{r}(k)~.
 \eeq
As expected, the curvaton field fluctuations are constant until
the curvaton starts to oscillate in the quadratic potential and
the field fluctuations begin to fall off as $a^{-3/2}$, just like the
background field $\sigma$.

\subsection{Perturbation amplitude at decay}

There are now two cases to investigate. The case with a non-vanishing
background value of the curvaton field, and the case of
vanishing background field. Let us first consider the more
interesting first case\footnote{A vanishing background field
leads to non-Gaussian perturbations, which are ruled out by
observations \cite{Lyth:2001nq,Lyth:2002my}.}. In this
case we can write
 \beq
\rho_{\sigma}=\frac{1}{2}\Omega^2m^2 \sigma^2~~\qquad
\delta\rho_{\sigma}(k)=m^2\Omega^2\sigma\delta\sigma(k)~,
 \eeq
where $\Omega=\exp(l\phi/2+mb/2)$. The curvaton density
perturbation becomes
 \beq
\delta\equiv \frac{\delta\rho_{\sigma}(k)}{\left<
\rho_{\sigma}\right>}=2\frac{\delta\sigma(k)}{\sigma}~.
 \eeq
So after the axion has started to oscillate in its potential we
get
 \beq \label{psspec3.1}
\frac{\delta\sigma(k)}{\sigma}=\frac{\tilde{A}}{C(\mu)}\frac{\delta\sigma_r(k)}{\sigma_r}
 \eeq
where we used that $\sigma_{osc}=\sigma_r$ and after the
curvaton has started to oscillate in the potential its amplitude
falls off like $a^{-3/2}$. It is interesting to note that even if
the curvaton field fluctuations depend on $\Omega$, the
fluctuations in the curvaton energy density itself does not. This
peculiarity has also been noted earlier, see for instance
\cite{Copeland:1998ie}. This also implies that the relation in
equation (\ref{psspec3.1}) implicitly depends on $\Omega$ through
$\tilde{A}$.

Finally
 \beq \label{pspec4}
\frac{k^{3/2}\delta\sigma(k)}{\sigma}
=k^{3/2}\frac{\tilde{A}}{C(\mu)}\frac{\delta\sigma_r(k)}{\sigma_r}
=\tilde{A}\frac{H_r}{\sigma_r}\left(\frac{k}{k_r}\right)^{3/2-\mu}~~,\qquad
\eta>\eta_{osc}~
 \eeq
which was obtained for the PBB curvaton model in \cite{Enqvist:2001zp,Bozza:2002fp} and for inflation in \cite{Lyth:2001nq} up to the numerical factor $\tilde{A}$.

For completeness we mention that, if the background field is
vanishing $(\sigma(k) = \delta \sigma(k))$, the power spectrum of
density perturbations is given by
 \bea
\int \frac{d^3 k}{(2\pi
k)^3}e^{i\vec{k}\cdot(\vec{x}-\vec{x}')}\mathcal{P}_{\rho}(k)&=&\left<\rho_x(\sigma)\rho_{x'}(\sigma)
\right>-\left<\rho_x(\sigma)\right>^2\nonumber\\
\Omega^4\frac{m^4}{4}(\left<\sigma_x^2\sigma_{x'}^2
\right>-\left<\sigma_x^2\right>^2)&=&\Omega^4\frac{m^4}{4}\int
\frac{d^3 k}{(2\pi
)^3}e^{i\vec{k}\cdot(\vec{x}-\vec{x}')}\int\frac{d^3
p}{(2\pi)^3}|\sigma_p|^2|\sigma_{k-p}|^2~.
 \eea
Thus,
 \beq
\mathcal{P}_{\rho}(k)=\Omega^4\frac{\tilde{A}^4}{16\pi}mH_1\left(\frac{k}{k_r}\right)^3
\left(\frac{k_r}{a}\right)^6\int\frac{dp}{p}\left(\frac{p}{k_r}\right)^4\frac{|k-p|}{k_r}n_pn_{k-p}
 \eeq
where $n_k=|k\eta_r|^{-2\mu-1}$. The integral was calculated in
\cite{Durrer:1998sv} and it implies the following power spectrum
 \beq \label{pspec3}
\frac{\mathcal{P}_{\rho}(k)}{\rho_c^2}=\frac{\Omega^4\tilde{A}^2B}{16\pi^2(\mu-1)}\frac{1}{9}\left(\frac{H_r}{m_p}\right)^4
\frac{m}{H_{r}}\left(\frac{H_r}{H}\right)^4\left(\frac{a}{a_r}\right)^6
\left(\frac{k}{k_r}\right)^{6-4\mu}
 \eeq
where
 \beq
B=
\frac{2^{4\mu-4}}{\sqrt{\pi}}\Gamma(2-2\mu)\Gamma(2\mu-3/4)\left[\cos2\pi(\mu-1)-1\right]~.
 \eeq
Note that also the spectrum in equation (\ref{pspec3}) is
constant in the matter dominated era. At times $\eta>\eta_{eq.}$,
one gets
 \beq
\frac{\mathcal{P}_{\rho}(k)}{\rho_c^2}=\frac{\Omega^4\tilde{A}^2B}{16\pi^2(\mu-1)}\frac{1}{9}\left(\frac{H_r}{m_p}\right)^4
\frac{m}{H_{eq.}} \left(\frac{k}{k_r}\right)^{6-4\mu}~.
 \eeq

\section{COBE normalization}

If the curvaton starts to dominate energy density only after is
has started to oscillate in the potential we must require
$\Omega\sigma_r<1$ in Planck units. Otherwise a short era of
inflation will arise \cite{Bozza:2002fp}. Normalizing to the COBE
observations, we must require
 \beq \label{bound1}
10^{-5}\simeq \mathcal{P}_{\zeta}^{1/2}(k_0)\simeq
\frac{k_0^{3/2}\delta\rho_r(k_0)}{\rho_r}
 \eeq
which implies that a flat spectrum of CMB perturbations is only
possible if \cite{Lyth:2001nq}
 \beq
H_r<10^{-5}m_p~.
 \eeq
This is more or less the same as the constrain on the Hubble scale 
during inflation obtained in \cite{Liddle:1993ch} from COBE observations. 
However for PBB the situation is more interesting. In the PBB scenario 
we expect $M_s\simeq H_r$ and a flat spectrum would require
 \beq \label{bound1.1}
M_s<10^{-5}m_p
 \eeq
as observed in \cite{Enqvist:2001zp,Bozza:2002fp}. This is
a very low value for the string scale. It looks like the
amplitude of the curvaton field fluctuations is too large to be in
agreement with the standard string scale and the COBE
normalization. However, there are several ways how one can
circumvent this problem. In \cite{Enqvist:2001zp} we suggested
that the curvaton is to be identified with an axion with a
periodic potential \footnote{One might fear that a periodic potential will
lead to formation of topological defects. However, if the potential has only 
one degenerate minimum i.e. if the anomaly factor is one ($N=1$), then 
the topological defects decay and do not pose a cosmological problem 
\cite{Kasuya:1996ns}.}, such that the periodic potential will damp
the density fluctuations to the right level. For a quadratic
curvaton potential it was suggested in \cite{Bozza:2002fp} that a
kink in the spectrum might also do the job of lowering the
curvaton field fluctuations to the right level.

There also exists another interesting possibility; the curvaton
might decay before it dominates the energy density\footnote{This
was first suggested by Lyth and Wands \cite{Lyth:2001nq}.}. The
curvature perturbation is then given by \cite{Lyth:2001nq}
 \beq \label{lyth}
\zeta=\frac{\rho_{\sigma}}{4\rho_{rad}+3\rho_{\sigma}}\frac{\delta\rho_{\sigma}}{\rho_{\sigma}}~.
 \eeq
If $\rho_{rad}>>\rho_{\sigma}$ at the decay, we get from 
equation (\ref{lyth}) and (\ref{psspec3.1})
 \beq \label{lyth1}
\zeta_d\simeq \frac{1}{2}\left(\frac{\rho_{\sigma}}{\rho_{rad}}\right)_d\frac{\delta\sigma_r(k)}{\sigma_r}
 \eeq
where $\rho_{rad}$ is the energy density of radiation and subscript $d$ 
denotes the point when the curvaton decays. Thus, if the PBB curvaton field 
decays when it only contributes $r=10^{-3}$ of the total energy density,
then by means of equation (\ref{lyth1}) we would get
$M_s=10^{-2}m_p$. This is within the current constrains on
non-Gaussianity from COBE data, which yield $r>6\times 10^{-4}$
(see \cite{Lyth:2002my} for details on the non-Gaussian aspects of
the curvature perturbations). In this case the curvaton can not
be the axion suggested in \cite{Enqvist:2001zp}, but must have a
stronger interaction with the photons than what an axion can
offer. This possibility is exciting since $r\simeq 10^{-3}$ can
already be ruled out or be confirmed by the MAP and PLANCK
satellites. Note, that it was recently shown by Lyth, Ungarelli and
Wands \cite{Lyth:2002my} that if $r<<1$ then CDM cannot be
created before the curvaton decays or by the curvaton decay
itself because large correlated or anti-correlated isocurvature
perturbations are produced.

In the next subsection we will see how an intermediate string
phase might also do the job of lowering the curvaton field
fluctuations to the right level in order to be in agreement with a
high string scale and the COBE normalization.

\subsection{Intermediate string phase}

Assume that there is an intermediate string phase
$\eta_s<\eta<\eta_r$ during which the curvaton field fluctuations
are frozen on super horizon scales. We parameterize our ignorance
about the string phase like in \cite{Brustein:1997wt}. We let
$z_s=a_r/a_s$ denote the ratio of the scale factor at the end of
the string phase and at the beginning of the string phase.
Likewise, we denote by $g_r=\exp(\phi(\eta_r)/2)$ and
$g_s=\exp(\phi(\eta_s)/2)$ the string coupling constant at
respectively the end and the beginning of the string phase. 
In order to have enough inflation in the pre-big bang era, 
we must assume $1<z_s<10^{20}$ \cite{Brustein:1997wt}. 
It is also natural to assume $g_r/g_s>1$.

Since it is $\delta\sigma(k)$ which is constant on superhorizon
scales and not the canonical normalized field $\psi_k$, we should match
$\delta\sigma(k)$ at $\eta_s$ and $\eta_r$. Using equation (\ref{pspec})
and (\ref{pspec1}) the matching leads to
 \beq \label{pspec5}
S_r\delta\sigma(k) = C(\mu)\frac{S_r}{S_s}\left(\frac{\eta_s}{\eta_r}\right)^{-\mu+1/2}\frac{1}{(2\eta_r\al)^{-1/2}}|k\eta_r|^{-\mu}y_2(b,x)~~,\qquad\eta>\eta_r~.
 \eeq
By comparing to equation (\ref{pspec1.2}) we see that the curvaton
fluctuation spectrum is multiplied by an additional factor
of\footnote{If we write $S_r/S_s=(k_r/k_s)^{-\delta}$, then
$\delta$ agrees with the parameterization of the break in the
spectrum discussed by Bozza $et$ $al.$ in \cite{Bozza:2002fp}.}
 \beq
\frac{S_r}{S_s}\left(\frac{\eta_s}{\eta_r}\right)^{-\mu+1/2} =
z_s^{-\mu+3/2}\left(\frac{H_r}{H_s}\right)^{-\mu+1/2}\left(\frac{g_r}{g_s}\right)^l\left(\frac{\be_r}{\be_s}\right)^m
 \eeq
where we defined $\be(\eta)\equiv \exp(b(\eta)/2)$. When
normalizing to the COBE observations, this leads to a non trivial
dependence on $l,m$.

As an example of a model in which the parameters $l,m$ are
non-trivial, we consider a non-linear sigma model in Einstein
gravity where the scalar fields parameterize a
$SL(3,\mathcal{R})/SO(3)$ coset \cite{Copeland:1997pg}. In
particular we consider the following lowest order action in 4-D
Einstein frame \cite{Lidsey:1999mc}
 \bea
S&=&\frac{1}{2\kappa^2}\int
d^4\sqrt{-g}\left[R-\frac{1}{2}(\nabla\phi)^2-\frac{1}{2}(\nabla b
)^2\right.\nonumber\\
&&\left.-\frac{1}{2}e^{\sqrt{3}b+\phi}\left(\nabla
\sigma_3\right)^2-\frac{1}{2}e^{-\sqrt{3}b+\phi}\left(\nabla
\sigma_2\right)^2-\frac{1}{2}e^{2\phi}\left(\nabla\sigma_1-\sigma_3\nabla\sigma_2\right)^2\right]~,
 \eea
where $\kappa^2=8\pi/m_p^2$, $\phi$ is the 4-D dilaton, $b$ is the
moduli, $\sigma_1$ is the NS-NS axion and $\sigma_2$, $\sigma_3$
are RR axions. This model has been discussed in great detail in
\cite{Copeland:1997pg,Copeland:1998ie,Lidsey:1999mc,Copeland:vi,Copeland:1997ug}. The action contains no potential terms,
but one expects them to be generated non-perturbatively. 
The perturbations of the axion with the smallest
tilt will dominate the energy density perturbation spectra.
It has been demonstrated that there are regions of parameter space where
either of the axions can be the dominating one and with a flat scale
invariant spectrum \cite{Copeland:1998ie}. We note that the conformal
factors for the three axion fields are \cite{Copeland:1998ie}
 \beq
\Omega_i^2=\left\{\begin{array}{ccc}
  e^{2\phi} & for & \sigma_1 \\
  e^{\phi-\sqrt{3}b} & for & \sigma_2 \\
  e^{\phi+\sqrt{3}b} & for & \sigma_3
\end{array}\right.
 \eeq
Curiously, for the third axion field $l=1$, $m=\sqrt{3}$. For a flat spectrum
with $\mu=3/2$, we obtain from equation (\ref{bound1})
 \beq \label{bound2}
10^{-5}>\frac{g_r}{g_s}\frac{H_s}{H_r}\left(\frac{\be_r}{\be_s}\right)^{\sqrt{3}}\frac{M_s}{m_p}
 \eeq
which, depending of the detailed dynamics of the string phase,
can relax the bound in equation (\ref{bound1.1}) if for
instance\footnote{It is assumed that the extra dimensions are
contracting in the PBB phase as well as in the intermediate string phase
such that $\be_s>>\be_r$.} $\be_s/\be_r\sim g_r/g_s>>1$, $H_s/H_r\lesssim 1$.
But if we like in \cite{Buonanno:1997zk} take the intermediate string phase to be a period 
of constant curvature, frozen internal dimensions and linearly growing
dilaton in the string frame, then in the Einstein frame the curvature scale will not be constant
but vary like the string coupling $H_s/H_r\approx g_s/g_r$, so even a very long 
string phase of this kind will not improve on the bound in equation (\ref{bound1.1}) for $\sigma_3$
and for the Kalb-Ramond axion $\sigma_1$ with $l=2$ it gets worse. 
A dual dilaton phase (also considered in \cite{Buonanno:1997zk}) with frozen internal 
dimensions and decreasing curvature scale in the string frame might even
be problematic since it would change the bound in equation (\ref{bound1.1}) in
the wrong direction. However, it is clear that an intermediate string phase
of accelerated expansion and increasing curvature scale 
(as considered in \cite{Gasperini:1998bm}) only have to
be very short in order to relax the bound in equation (\ref{bound1.1}), with
the curvature scale growing only two orders of magnitude.  
If the internal dimensions are contracting, we can obtain a high string scale 
even with a constant curvature scale during the intermediate string phase (in string frame)
with $\sigma_3$. If the internal dimensions contract fast enough the $\sigma_3$ 
field might even work as a curvaton if we have a decelerating dual dilaton phase.

The discussion above is of course just an example.
Generally we expect as many as 15 different axion fields in a
$SL(6,\mathcal{R})$ invariant model, which is the maximal
invariance for toroidal compactification from 10 to 4 dimensions,
all with different conformal factors \cite{Bridgman:2000kk}. 
Note that in the simplest case of sudden transition without an intermediate 
string phase, all the different axion fields have the same amplitude for 
modes crossing the Hubble scale at the start of the post big bang era 
$k\approx k_r$. On larger scales the amplitudes are fixed by the different
spectral tilts and the field with the smallest tilt yields the largest 
density perturbations. The increased symmetry group will enlarge the part of
parameter space where the spectral tilt of the dominating axion is close to flat
and where $l,m$ has non trivial values. In the maximal symmetric case, 
the 14 other axions will have a more blue spectrum and only contribute 
fractionally to the density perturbations. It might be possible that 
some of the fields are stable and can act as cold dark matter, leading also
to an isocurvature component in the density fluctuations. 
However, such scenarios depends on a more detailed understanding of how 
the non-perturbative potentials are generated. 

In this light it might be possible to obtain a flat spectrum in 
the PBB curvaton scenario with the standard string scale, 
but a more detailed understanding especially
of the nature of the graceful exit is still needed.

\subsection{Bounds on mass and reheat temperature}

From requiring that the energy density contributed by the
curvaton at the beginning of the radiation dominated epoch is
less than critical, we obtain an upper bound on the curvaton mass
and the reheat temperature. From
 \beq
\rho_{\sigma}<\rho_{c}~~,\qquad \eta=\eta_r
 \eeq
we have
 \beq
\frac{1}{2}\Omega^2m^2\sigma_r^2<3m_p^2H_r^2
 \eeq
which leads to
 \beq \label{bound4}
10^{-5}\simeq \mathcal{P}_{\zeta}^{1/2}\simeq
\frac{k^{3/2}\delta\rho_r}{\rho_r}>\frac{m}{m_p}~.
 \eeq
This is consistent also with the bound obtained in "the simplest
curvaton model" of Bartolo and Liddle \cite{Bartolo:2002vf}.

To get an upper bound on the reheat temperature, let us assume that
the curvaton interaction is suppressed by some mass-scale $M$ and
define $R=M/m_p$. Then the lifetime of the curvaton is given by
 \beq
\tau=M^2/m^3
 \eeq
and the reheat temperature is approximately given by
 \beq
T_{RH}\simeq \frac{m_p^{1/2}}{\tau^{1/2}}~.
 \eeq
Combined with our bound on the curvaton mass in equation (\ref{bound4}),
this leads to
 \beq
T_{RH}<\frac{1}{R}10^{-7}m_p~.
 \eeq
The reheat temperature is thus likely to be low enough to avoid
excessive production of gravitinos and monopoles, as was also the
case in the explicit example in \cite{Enqvist:2001zp}, provided that
$R\simeq 1$, which means that the curvaton decays only gravitationally.

\section{Summary}

We have followed the dynamics of the curvaton with a quadratic
potential till after the perturbation modes have become
non-relativistic. We found  explicitly, that the spectral tilt of
the curvaton stays the same throughout the dynamic evolution as was
suggested in \cite{Lyth:2001nq}. The case is similar to that of
the "seed mechanism" of the PBB scenario \cite{Durrer:1998sv,Gasperini:1998bm}.

The inflationary scale is constrained by observations to less
than five orders of magnitude smaller that the Planck scale, which
is in agreement with also the constrain from the curvaton
mechanism by normalizing to the COBE observations
\cite{Lyth:2001nq}. However, in the PBB scenario the curvature
scale at the beginning of the first radiation dominated epoch is
supposed to be near the string scale. This leads to an
inconsistency, when requiring both a flat spectrum and
normalization to the COBE observations \cite{Enqvist:2001zp,Bozza:2002fp}. 
It has been suggested before that either a periodic
potential \cite{Enqvist:2001zp} or a break in the spectrum
\cite{Bozza:2002fp} can solve this problem. If the curvaton decays
faster it could also help in this direction. This an interesting
possibility which could be verified or excluded already by observations 
from the MAP and PLANCK satellites and which leads to severe theoretical
constrains on the CDM due to the possible large isocurvature modes
\cite{Lyth:2002my}. But, depending of the specific model, one can
get potential problems with a too high reheat temperature 
if the curvaton couples too strongly to other fields.

It should be noted that if the string scale is of order the GUT
scale, then the perturbations have naturally the amplitude required by 
observations. Also, with a quadratic potential we obtain a slightly
blue spectrum for the curvaton, but with the periodic potential considered
in \cite{Enqvist:2001zp} it is also possible to have a slightly red spectrum even
with a high string scale. In any case there is still many things to be
understood. To obtain more precise predictions from the PBB curvaton
scenario, we need to understand more about the details of the non
perturbative potential of the axion and especially we will need to
know more about the nature of the graceful exit. 

Finally, we investigated the conditions under which an
intermediate string phase leading can lead to the observed level of
density fluctuations with a high string scale. The intermediate 
string phase is supposed to facilitate the graceful exit and various 
types of such phases has been considered in the literature. We found
that if the curvature scale is frozen during this phase the Kalb-Ramond 
axion has wrong coupling, while within a $SL(3,\mathcal{R})$
non-linear sigma model the $\sigma_3$ field due to its coupling to the
contracting internal dimensions is a consistent candidate as a curvaton.
If the intermediate string phase is a phase of accelerated expansion
even the Kalb-Ramond axion might be consistent with a high string scale,
while a dual dilaton phase of decelerated expansion could be a problem for
the Kalb-Ramond axion leading to a small string scale. In any case, if the 
internal dimensions contract fast enough, the $\sigma_3$ remains a consistent
candidate.

We showed that within a $SL(3,\mathcal{R})$
non-linear sigma model one of the three axions has the right
coupling to the dilaton to yield a flat spectrum even with a high
string scale, as long as the intermediate phase is not that of a 
decelerated expansion of the external dimensions and frozen internal dimensions.

\begin{center}
\textbf{Acknowledgment}
\end{center}

\noindent I would like to thank prof. Kari Enqvist for motivating
discussions and comments on the manuscript.

\end{document}